\documentclass[sigconf]{acmart}

\usepackage{comment}
\usepackage{xspace}
\usepackage{enumitem}
\usepackage{amsfonts}

\AtBeginDocument{%
  \providecommand\BibTeX{{%
    \normalfont B\kern-0.5em{\scshape i\kern-0.25em b}\kern-0.8em\TeX}}}

\acmConference[CIKM '20]{
    29th ACM International Conference on Information and Knowledge Management
}{October 19--23, 2020}{Galway, Ireland}

\definecolor{orange}{RGB}{255,119,0}
\definecolor{red}{RGB}{220,0,0}
\definecolor{agreen}{RGB}{74, 198, 148}
\definecolor{purple}{RGB}{158, 62, 177}
\definecolor{darkpurple}{RGB}{170, 70, 210}
\definecolor{aqua}{RGB}{87, 180, 181}
\definecolor{lightblue}{RGB}{72, 123, 232}
\definecolor{hotpink}{RGB}{255, 83, 115}
\definecolor{linkColor}{RGB}{6,125,233}

\newcommand{\tool}[0]{\textsc{PeopleMap}\xspace{}}

\begin{document}

\title{\tool{}: Visualization Tool for Mapping Out Researchers using Natural Language Processing}

\author{Jon Saad-Falcon, Omar Shaikh, Zijie J. Wang, Austin P. Wright, Sasha Richardson, Duen Horng Chau}
 \affiliation{Georgia Institute of Technology, Atlanta, GA, USA}
 \email{{jonsaadfalcon, oshaikh, jayw, apwright, polo}@gatech.edu}
 \affiliation{Fayetteville State University, Fayetteville, NC, USA}
 \email{srichardson@broncos.uncfsu.edu}

\renewcommand{\shortauthors}{Saad-Falcon, et al.}

\begin{abstract}

Discovering research expertise at institutions can be a difficult task. Manually curated university directories easily become out of date and often 
lack the information necessary for understanding a researcher's interests and past work, making it harder to explore the diversity of research at an institution  and identify research talents. This results in lost opportunities for both internal and external entities to discover new connections and nurture research collaboration.

To solve this problem, we have developed \tool{}, the first interactive, open-source, web-based tool that visually ``maps out'' researchers based on their research interests and publications by leveraging embeddings generated by
natural language processing (NLP) techniques. \tool{} provides a new engaging way for institutions to summarize their research talents and for people to discover new connections. 
The platform is developed with ease-of-use and sustainability in mind. 
Using only researchers' Google Scholar profiles as input, \tool{} can be readily adopted by any institution using its publicly-accessible repository and detailed documentation.

\end{abstract}



\begin{teaserfigure}
\centering
  \includegraphics[width=0.9\textwidth]{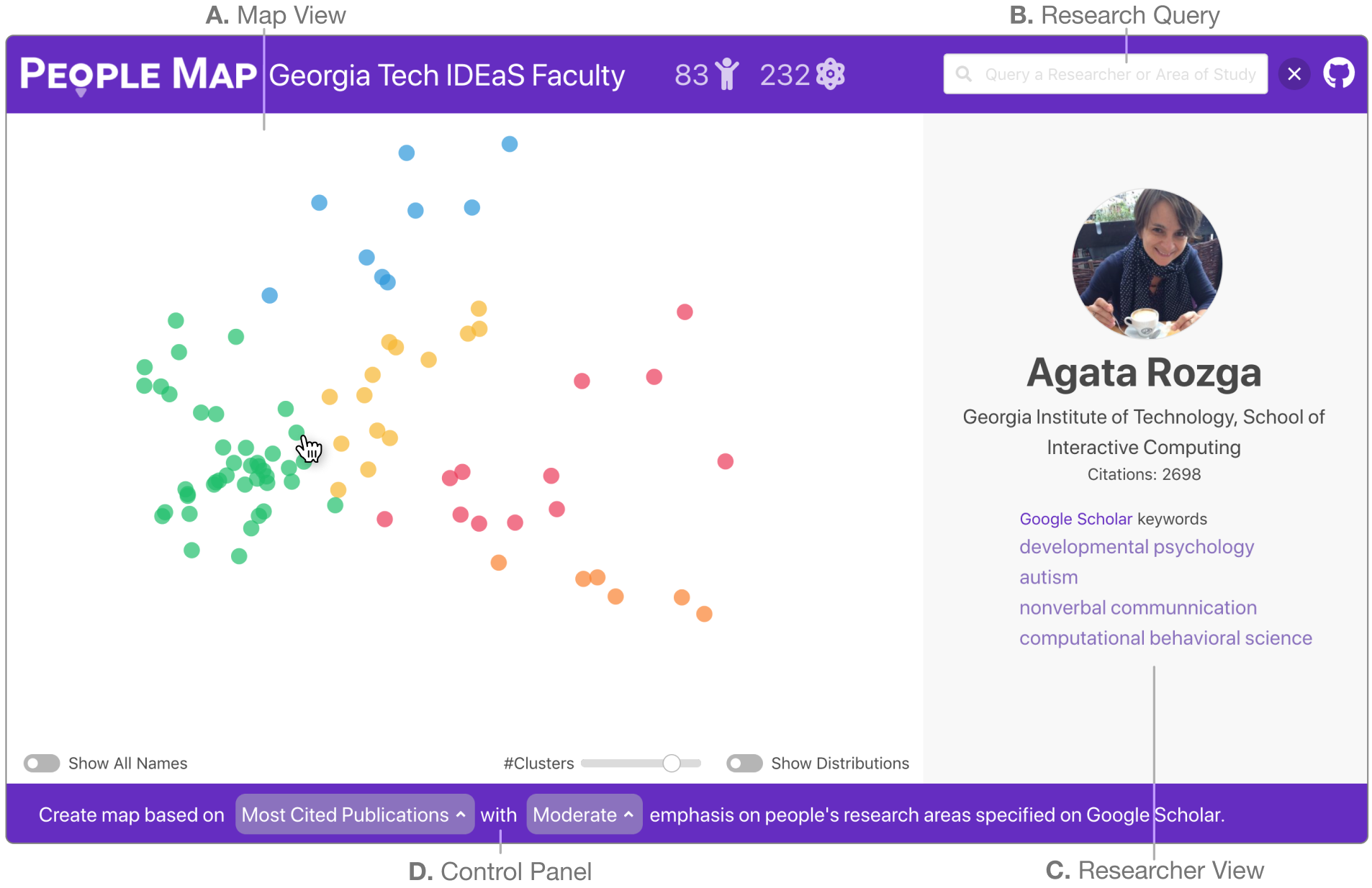}
  \caption{
  \textmd{\tool{} visually maps out researchers based on their research interests and publications.
  Here, a \tool{} user is exploring the 
  research topics
  of the faculty members at the \textit{Institute of Data Engineering and Science} (IDEaS) at Georgia Tech (\textcolor{linkColor}{\url{https://poloclub.github.io/people-map/ideas/})}} 
  A. \textmd{\textbf{Map View} visualizes the embedding of researchers generated using their research topics and publication data, with each dot representing a researcher.}
  B. \textmd{\textbf{Research Query} allows users to search for researchers and query areas of study,
  allowing the user to both locate specific individuals and see the researchers most associated with a queried field in the Map View.} 
  C. \textmd{\textbf{Researcher View} shows the detailed information (e.g., affiliation, citations, interests) of a researcher highlighted in Map View.
  }
  D. \textmd{\textbf{Control Panel} allows users to adjust the hyperparameters of the \textit{Map View} visualization (e.g., show research names and cluster information).}
  }
  \label{fig:teaser}
\end{teaserfigure}

\maketitle

\section{Introduction}
The task of discovering research expertise at institutions pursuing scientific research can be a difficult task. For example, it is often challenging for companies or research laboratories to pinpoint researchers at an institution to collaborate with for the problems they wish to solve \cite{belkhodja2007triple}; this can lead to methods such as sending scores of emails to faculty, hoping that eventually the appropriate individual will be found. However, this method often leads to lost opportunities. Furthermore, within institutions themselves, the task of planning major research efforts often involves the same tactics when it comes to finding appropriate people. 

Both of these situations share a similar problem: the lack of an easily accessible, up-to-date record with relevant information for user queries. At the moment, most institutions have manually curated directories, which hold each individual's department, position, and contact information. However, when trying to learn more about a researcher these directories can sometimes provide inaccurate or incomplete information. For instance, a full professor's latest research interests may be very different from those at the time that faculty member joined the institution years ago.
Additionally, the information typically lacks important details about their specific fields of interest as well as their publications. Therefore, these directories can be less helpful for the companies, laboratories, and government agencies that seek to pursue business or fund projects at a specific university. This hinders the ability of both external and internal individuals to pinpoint research talents, understand the scope of research activities at an institution, and discover new connections.

\section{Contributions}
To solve this common challenge faced by research institutions, our work makes the following contributions:

\begin{enumerate}[itemsep=2mm, topsep=2mm, parsep=1mm, leftmargin=5mm]

    \item \textbf{\tool{}}, an interactive tool that ``maps out'' researchers based on their research interests and publications by leveraging embeddings generated by natural language processing (NLP) techniques. 
    \tool{} contributes as:
    
    \begin{itemize}
    \item \textbf{The first visualization dedicated to helping users explore researcher embeddings}; while there has been research that develops methods to recommend research papers and publication venues \cite{beel2016paper, medvet2014publication, beel2017towards, alhoori2017recommendation, kuccuktuncc2013theadvisor}, less work focuses on developing usable easy-to-access tools for users to interactively explore researcher datasets. \tool{} fills this research gap and seeks to improve the interpretability and explorability of researcher datasets.

    \item \textbf{An open-source, sustainable web application for the community} that can be easily accessed via web browsers
    and implemented as a web-based application. 
    \tool{} is registered under the permissive MIT license, and its code repository is available at \textcolor{linkColor}{\url{https://github.com/poloclub/people-map}}.
    Besides the \tool{} visualization, 
    it also provides a series of data collection and preprocessing tools that allows users to create a researcher dataset from any list of researchers found on Google Scholar. Additionally, it includes a step-by-step documentation guide (\textcolor{linkColor}{\url{https://app.gitbook.com/@poloclub/s/people-map/}}) that covers every step of the process from downloading the repository to launching the \tool{} platform (Section \ref{sec:usages}).
    With the combined data collection resources and \tool{} visualization, the tool provides an \textbf{automated solution for researcher interest summarization and discovery}, which simplifies the exploration of the work of scientific researchers (Section \ref{sec:peoplemap}).
    \end{itemize}
    
    \item \textbf{\tool{} Usage Scenarios and Deployment} As a first real-world use case of \tool{}, we have successfully implemented \tool{} for \textit{The Institute for Data Engineering and Science}(IDEaS), a major cross-campus research entity at Georgia Tech  (\textcolor{linkColor}{\url{http://ideas.gatech.edu/}}) whose members include faculty from across colleges and departments on campus. Preliminary feedback from IDEaS' leadership team has been positive; they are very excited about \tool{}'s interactivity and the way that this tool can be easily updated for new members. The live \tool{} for IDEaS can be found at \textcolor{linkColor}{\url{https://poloclub.github.io/people-map/ideas/}}. 
    
    To demonstrate the easy application of \tool{} to a different organization's members, we also implemented \tool{} for the \textit{Center of Machine Learning} at Georgia Tech (\textcolor{linkColor}{\url{https://ml.gatech.edu/}}), another major cross-campus entity. 
    The live \tool{} for the Center of Machine Learning can be found at: \textcolor{linkColor}{\url{https://poloclub.github.io/people-map/ml/}}.
    We also provide an additional usage scenario to highlight how a potential user could implement and use \tool{}.
  
\end{enumerate}

\section{Introducing \tool{}}
\label{sec:peoplemap}

\tool{} is an open-source, web-browser-based visualization tool that maps out researchers using NLP techniques, allowing users to explore all the different information extracted from researchers' profiles using textual embeddings. It can determine the possible groupings of similarly-interested researchers, represent how researchers align with specified fields of study, and reveal potential Gaussian distributions describing the research topics present in the dataset.
\tool{}'s user interface consists of four major components:
(1) \textbf{Map View} (\autoref{fig:teaser}A) visualizes the 
research topic similarities among researchers;
(2) \textbf{Research Query} (\autoref{fig:teaser}B) allows users to search for researchers and query areas of study;
(3) \textbf{Researcher View} (\autoref{fig:teaser}C), which shows the detailed information of the researcher hovered over by the user (e.g., affiliation, citations, interests);
(4) \textbf{Control Panel} (\autoref{fig:teaser}D) allows users to adjust the hyperparameters of the Map View visualization. 
Next, we describe each component in more detail.

\subsection{Mapping Out Researcher Interests}

The Map View of \tool{} (\autoref{fig:teaser}A) is a visualization of embeddings representing the researchers in the selected dataset. Within the Map View, each dot represents a researcher and their corresponding embedding projected into a two-dimensional space. With the researcher data extracted from Google Scholar, these embeddings were created using term frequency–inverse document frequency (TFIDF) \cite{jones1972statistical} matrices and principal component analysis (PCA) \cite{wold1987principal}, which is discussed in greater detail in the following sections:

\subsubsection{\textbf{Collecting Google Scholar data for each researcher}}

Generating a \tool{} visualization requires only public data that anyone can access.
We collect each researcher's public information from Google Scholar, which includes the researcher's profile, publications, and research interests using a Python-based module called \textit{scholarly} (\textcolor{linkColor}{\url{https://github.com/scholarly-python-package/scholarly}}). The specific information included are:

\begin{itemize}

\item Google Scholar profile URL
\item Top 50 most cited publications (titles, abstracts, and years of publication)
\item Top 50 most recent publications (titles, abstracts, and years of publication)
\item Google Scholar profile keywords
\item Citation count
\item Institution affiliation
\item Google Scholar profile photo

\end{itemize}

\noindent
\tool{} formats and stores all researcher data in a CSV file,
one column for each category of information listed above.

\subsubsection{\textbf{Researcher Embeddings}}
\label{sec:wordembedding}
    
Using the publication data extracted from Google Scholar, the title and abstracts of each researcher's publications are first concatenated together to create a combined document for each researcher. 
Additionally, Google Scholar keywords of each researcher can also be concatenated into their respective combined documents.
After their creation, in order to normalize and prepare them for analysis, these combined documents are:

\begin{enumerate}
    \item \text Removed words with non-English alphabet characters to restrict the bounds of analysis
    \item \text Eliminated words with fewer than two characters in length to mitigate noise in the data
    \item \text Lowercased words to simplify capitalization
    \item \text Cleaned of HTML tags
    \item \text Cleaned of stop-words
    \item \text Stemmed words to simplify syntax
\end{enumerate}{}

Once the documents have been normalized, they are then converted into researcher embeddings representing each individual researcher through the use of the TFIDF technique. 
This technique takes into account both the occurrence of each word within a researcher's publications and its frequency. Furthermore, it provides us a quantitative method by which we can ignore common words shared by most, if not all, of the researchers, while measuring specific ``important'' or ``characteristic'' words that differentiate researchers \cite{jones1972statistical}. 
Each researcher's embedding is a column in a TFIDF matrix, with each row representing the respective term values for a specific word in each researcher's embedding. 
The following equation represents the combination of $n$ total researcher embeddings, 
each individually represented as vectors $v$, 
to create the combined TFIDF matrix $\mathbf{R}$:  \[ [v_1, v_2, ..., v_n] = \mathbf{R}\]

With the researcher embeddings in the TFIDF matrix, it is necessary to first reduce the dimensionality of the embeddings, which are vectors in a several-thousand dimensional space, so that they can be visualized.
To achieve this, principle component analysis (PCA) is used to assist in feature extraction and elimination, simplifying the researcher embeddings into vectors within a two-dimensional space that can be visualized in the Map View (\autoref{fig:teaser}A). 

We chose PCA as a starting embedding technique, because \tool{} is one of the first tools for interactively mapping out researchers. 
Our primary goal is to create a platform that improves the explorability and interpretability of researcher datasets. 
While there are many potential embedding techniques for the textual data of researchers, we aimed to start with more classic embeddings that could provide adjustable parameters for the platform. 
We purposefully used PCA over other potential visualization techniques, such as UMAP \cite{mcinnes2018umap} or t-SNE \cite{maaten2008visualizing}, 
because they tend to find structure within the noise of a dataset with small sample sizes compared to the dimensionality of the data, 
while PCA is well justified as a linear model for such datasets \cite{mcinnes2018umap}. 
Thus, we use PCA since it fits the constraints of our researcher dataset and allows us to still find emergent patterns among the researcher embeddings.
In the future, we endeavor to improve the complexity of our embeddings by exploring several potential embedding techniques.

\subsection{Querying Researchers and Areas of Study}
    \label{sec:ResearchQuery}
    
\begin{figure}[t]
\centering
    \includegraphics[width=\linewidth]{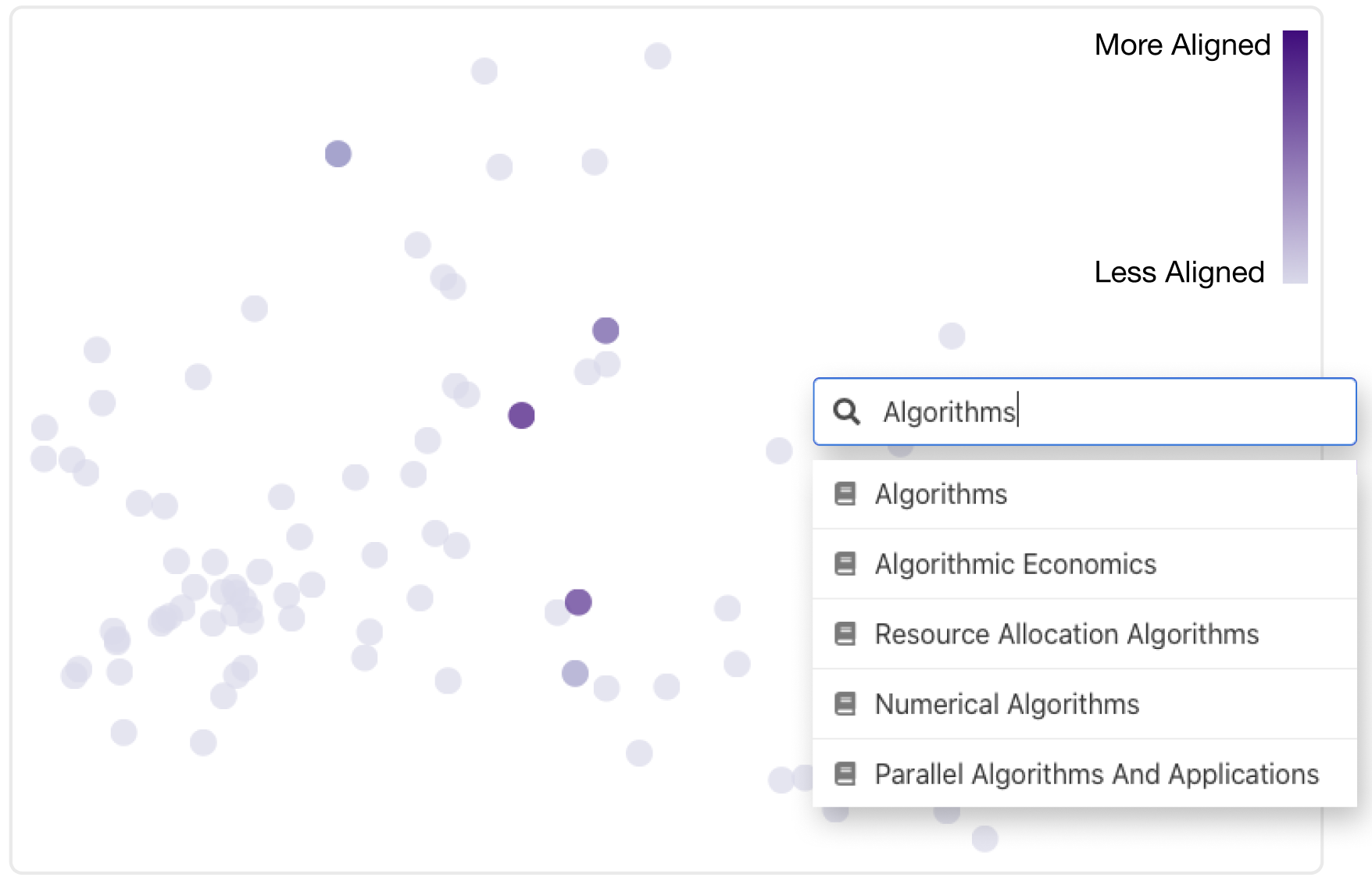}
    \caption{Research Query component and query results displayed in Map View. 
    \textmd{Researchers are colored based on how well they align with the query (in this example, ``Algorithms'' is the query research topic).  
    Darker means more aligned.
    }}
    \label{fig:ResearchQuery}
\end{figure}
   
	The Research Query component (\autoref{fig:ResearchQuery}) allows the user to both locate specific researchers, as well as see which researchers are aligned with each of the Google Scholar keywords collected from the researcher dataset. 
	When the user searches for a researcher, \tool{}{} highlights the researcher's representation in Map View by enlarging the dot's radius and outlining it; \tool{} also displays the researcher's Google Scholar profile information in the Researcher View. 
	When calculating a researcher's alignment with a selected Google Scholar keyword, \tool{} uses similarity analysis between researcher embeddings and topic embeddings, which is discussed in the next section.
	
	\subsubsection{\textbf{Similarity Analysis}}

    The TFIDF researcher embeddings used for the Map View component are also used to calculate the similarity between a researcher and a specified topic. 
    For example, if a user wants to see which researchers frequently use a specific term prominently throughout their work, it is possible to use their researcher embeddings to find which ones use the term most often compared to their overall writing. 
    To calculate this, the specified topic (e.g. ``natural language processing'') is first converted into a TFIDF embedding using the same process that is outlined for the researcher's publications in (Section \ref{sec:wordembedding}). 
    Then, the cosine similarity between the specified topic embedding and each of the researcher embeddings in the TFIDF matrix is calculated, which indicates the similarity between the two vectors: the higher the value, the greater the similarity~\cite{ramos2003using}. 
    The following equation represents the cosine similarity between the specified topic embedding, represented as the vector $q$, and the current researcher embedding, represented as the vector $v$, to produce the resulting similarity score, represented as $s$: \[ \frac{q \cdot v}{\left \| q \right \| \times \left \| v \right \|} = s\]
    
    By performing cosine similarity calculations between the specified topic embedding and the researcher embeddings in the TFIDF matrix, the top similarity scores can be used to find the researchers that most align with the specified topic. 
    These researchers are, in turn, highlighted in the Map View when the specified topic is queried in the Research Query component (\autoref{fig:ResearchQuery} shows an example query).
    Researchers are colored based on how well they align with the query. Darker indicates more aligned. The Research Query tool, together with the color gradient visualizing the query results, help users better understand the scope of research relevance among the researchers.
    The researchers more prominently highlighted are those who tend to use the query term proportionally more than their peers in the dataset. 
    This can serve as a reference to begin inquiries into the individual's research rather than serve as a full assessment of their contributions to that research topic.

\subsection{\textbf{Clustering Researchers}}
\label{sec:clusteringtechniques}

To help users more easily identify groups of related researchers, the Map View (\autoref{fig:teaser}A) colors the researcher dots to indicate clusters of associated researchers. 
The intention of this coloration is not to create strictly-defined groups of researchers.
Rather, we want to help users visualize the scope of shared interests 
among researchers. 
To assign these colorings, we use Gaussian mixture modeling, which will be explained in greater detail in the following section.

\subsubsection{\textbf{Assigning Colors}} Previously, we used PCA to reduce the dimensionality of the researcher embeddings, projecting them into a two-dimensional space for visualization (Section \ref{sec:wordembedding}). 
This dimension reduction of the researcher embeddings is also necessary for clustering techniques to be performed. 
In the researcher dataset for the IDEaS faculty at Georgia Tech (which is visualized in Figure \ref{fig:teaser}), the researcher embeddings have over 11,000 dimensions, with each dimension representing a word in the vast vocabulary shared by the researcher dataset; 
however, there are only 83 datapoints. Thus, considering the complexity of the data, it is necessary to simplify the dimensionality of the data before performing clustering~\cite{bellman2015adaptive}.

Therefore, using the newly-reduced researcher vectors created using PCA, the total set of researcher vectors is analyzed using Gaussian mixture modeling. 
Using this technique, the overall distribution of researcher vectors is categorized into several different Gaussian distributions (visualized distributions in \autoref{fig:Distribution}). 
These distributions are meant to assist the user in their understanding of the different topics within the researcher dataset and how these topics are shared among different groups.

Once these researcher vectors are clustered using Gaussian mixture modeling, they are visualized within the Map View component of \tool{} and colored according to their designated Gaussian distribution, with each distribution being assigned a unique color.  Researcher dots that are close together tend to reflect a similarity in research pursuits between the two researchers; increased distance between researcher dots reflects the opposite. Using distance and coloring of a research embedding as metrics for gauging similarity, the user can better understand the relationship between each of the researchers as well as the diversity of topics in the Map View (\autoref{fig:Distribution}).

\subsection{Calibrating Exploration}
\label{sec:calibratingexploration}

To change the settings of the Map View, the user can use the control panel at the bottom of the visualization (Figure~\ref{fig:teaser}D) to manipulate the Map View and investigate the relationships and information presented by the dataset. The following settings assist in the exploratory process of the researcher dataset, allowing the user to explore the impact of different variables on the overall visualization and patterns among the researchers. 

\begin{itemize}

\item \textbf{Show Distributions}: In order to help the user better understand how each cluster of researchers is formed, this toggle permits the user to see the Gaussian distributions calculated by the Gaussian mixture model (discussed in  Section~\ref{sec:clusteringtechniques}). Each distribution is colored differently according to the dots within it. Additionally, each distribution visualizes the space covered by three standard deviations of the distribution along each of its axes. 

\begin{figure}[t]
    \centering
    \includegraphics[width=\linewidth]{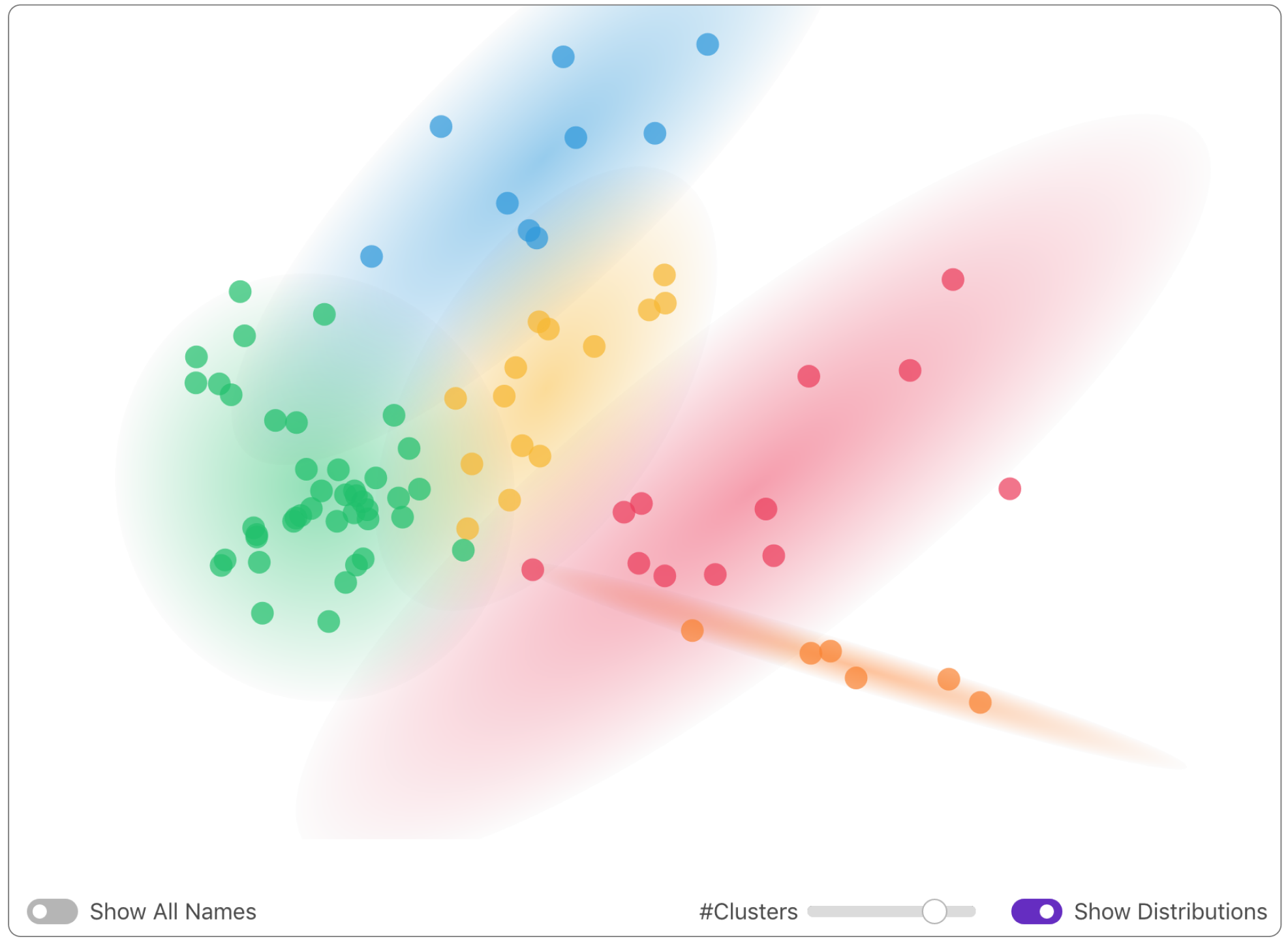}
    \caption{Map View component with cluster distributions displayed. \textmd{Each dot is an embedding of a researcher’s combined publications; the color is assigned by their membership in a cluster of a Gaussian mixture model, which are displayed as elliptical distributions. 
    Proximity of dots indicates similarity of research interests while remoteness indicates disparity. }}
    \label{fig:Distribution}
\end{figure}

\item \textbf{\#Clusters}: To assist the user in their exploration of the researcher embedding clusters, this slider allows the user to adjust the number of Gaussian distributions generated by the Gaussian mixture model algorithm (discussed in Section~\ref{sec:clusteringtechniques}). The slider itself does not change the embeddings of the researchers. 
By increasing the number of clusters, 
the Gaussian distributions generated becomes increasingly tight.
Likewise, by decreasing the number of clusters, the Gaussian distributions become more expansive but also decrease in tightness.

\item \textbf{Show All Names}: To help users find specific researchers and recognize individuals in different clusters, this toggle displays the names of researchers alongside their respective dot within the Map View. It can be used to find researchers without hovering over each dot individually.

\item \textbf{Keywords Emphasis}: This drop-down allows the user to adjust the emphasis that is placed on a researcher's Google Keywords, compared to their titles and abstracts, when generating their TFIDF embedding (Section \ref{sec:wordembedding}). By increasing the emphasis, more multiples of a researcher's keyword are concatenated into their original combined document that is used to generate their TFIDF embedding. By decreasing the emphasis, less multiples of a researcher's keywords are concatenated into their original combined document. 
The purpose of this drop-down is to increase or decrease the weight placed on a researcher's self-identified topics of study when calculating their position in the visualization, allowing the user to better understand the characteristics of each researcher's fields of interest.

\item \textbf{Publication Set}: This drop-down allows the user to select which publications they would like to use for the Map View: 
The default option will use a researcher's 50 most cited publications to characterize their research, 
while the other option will use a researcher's more recent 50 publications in their characterization. 
These options allow users to explore the researcher dataset from two different angles of what a researcher may be more known for and what they are currently working on. 

\item \textbf{Researcher details on demand}: To see more information about a researcher, users can hover over the researcher's dot in the Map View (\autoref{fig:teaser}A), which will display in the Researcher View (\autoref{fig:teaser}C) the researcher's:

\begin{itemize}
\item Name
\item Affiliation
\item Google Scholar Profile Keywords
\item Keywords
\item Total Citation Count
\item Google Scholar Profile Link
\item Google Scholar Profile Photo
\end{itemize} 
\end{itemize}

\section{Usage and Access of \tool{}}
\label{sec:usages}

\subsection{\tool{} Code Repository and Documentation}
\label{sec:PeopleMapRepository}

	The source code for \tool{} is available at \textcolor{linkColor}{\url{https://github.com/poloclub/people-map}}; it is registered under the permissive MIT license, making it available to anyone.
	It includes the \tool{} visualization as well as the data collection and processing code for developing a new researcher dataset which can be loaded into the platform. Furthermore, our documentation provides concrete tutorial steps for users to follow, so that new users with beginner's experience in Python and Javascript may also easily set up the tool. It walks a new user through the initial steps of collecting data from Google Scholar to the final stages of setting up the \tool{} platform on their computer.
    
    In addition to the source code, we provide two live demos of \tool{} that
    allow anybody to explore and become familiar with the \tool{} platform. 
    The first demo analyzes the publications of the faculty in Georgia Tech's \textit{Center of Machine Learning} (\textcolor{linkColor}{\url{https://poloclub.github.io/people-map/ml/}}), 
    while the second demo analyzes the publications of the faculty at the Institute for Data Engineering and Science (IDEaS), also at Georgia Tech  (\textcolor{linkColor}{\url{https://poloclub.github.io/people-map/ideas/}}). 

The corresponding datasets for these two faculty groups are available alongside the source code of the Github page: \textcolor{linkColor}{\url{https://github.com/poloclub/people-map}}.

\subsection{Example Usage Scenario}

James is an academic director at a university, looking to develop a new project centered around the study of black holes. He is looking for potential colleagues at his university with whom he can begin working on this new project. While he does have some current connections with professors at his university, he would like to explore the diversity of researchers at his university by using \tool{}. 

To start, James clones the \tool{} repository and begins following the steps of the documentation. Next, he goes to the university directory and gathers the Google Scholar profile names of all of the relevant researchers. Using tools included in the repository, he gathers their relevant publication information, processes the text, and generates the data files for the \tool{} platform.

With \tool{} fully set up, James begins exploring the researcher dataset with all the tools explained in Section \ref{sec:peoplemap}. First, he uses the \textit{Publication Set} drop-down and selects \textit{Most Recent Publications} since he wants to find researchers currently focusing on studying black holes. Next, James clicks the \textit{Research Query} component (Section \ref{sec:ResearchQuery}) and types ``black holes'', searching to see the researchers most closely aligned with the topic. The tool then highlights the top-five researchers associated with the topic. From this initial search, he discovers several individuals he did not know from his previous correspondence and decides to look a little deeper.

Using this information, James proceeds to use the \textit{Researcher View} component (\autoref{fig:teaser}C) to identify the researchers, clicking on their Google Scholar profile links to see some of their published work. 
However, before ending his search, he would like to see some of the other researchers that are in close proximity to the ones already selected. 
Using the \textit{Keywords Emphasis} drop-down, he tries different choices of keywords to see the groups of researchers that emerge near the previously identified researchers, using the \textit{Show All Names} toggle to take note of other researchers that are frequently associated with the ones found using the \textit{Research Query} component. 
With this wide array of researchers, James is confident he has gathered all the potential collaborators and proceeds to use their Google Scholar profiles found in the \textit{Researcher View} component, as well as other resources, to gauge which ones would be the best fit for the project.

\section{Predicted Impact}

\subsection{Enhanced and Enabled CIKM Research Activities}

\tool{} aims to facilitate several different CIKM research areas. As a tool for the visualization and exploration of researcher datasets, \tool{} seeks to assist in the \textbf{data presentation} of research fields of interest and researcher information. Furthermore, \tool{} can provide functionalities for \textbf{users and interfaces for information and data systems} by increasing the interactivity and explorability of researcher datasets through the \tool{} platform and its functionalities. By assisting in both of these CIKM research areas, \tool{} offers a new platform for public and private organizations to both explore the interests of their members and summarize the fields of study their members pursue.

\subsection{Scaling the Impact of \tool{}}
\label{sec:scaling}

\noindent
\textbf{\tool{} for research entities.}
\tool{} could transform how research talents at research institutions may be summarized and discovered by both internal and external collaborators. 
At Georgia Tech, we have successfully developed \tool{} for two major research entities: IDEaS and the Center for Machine Learning. 
The leadership of IDEaS are very excited about this tool, especially the interactivity and explorability that it provides for researcher datasets as well as the ease with which it can be updated for new members. While we used the tool for faculty datasets in IDEaS and the Center of Machine Learning in Section~\ref{sec:PeopleMapRepository}, it could be applied to the entirety of the College of Computing or even Georgia Tech as a whole. The scope of the researchers included is a matter of preference for the group seeking to implement \tool{}.

\medskip

\noindent
\textbf{\tool{} for larger entities.}
Using the data-collecting and processing tools that are part of the \tool{} repository, it is possible to expand the platform to other researcher datasets, as long as these researchers have Google Scholar profiles with their associated publications listed. 
The \tool{} for IDEaS visualizes 83 researchers.
However, it is possible to have significantly more researchers than this amount; the limiting factor for the total count is essentially the size of the Map View visualization. 
As more researchers are added, the higher number of dots can lead to greater visual complexity in the visualization, potentially causing ``overplotting'' as it becomes harder to distinguish between each of the dots and locate specific individuals using either the \textit{Show All Names} toggle or the Researcher View component. 
Additionally, the researcher information within \tool{} does not update automatically when researchers' Google Scholar profiles update.
\tool{} users would need to re-run the data collection and processing step to refresh \tool{}.

\medskip

\noindent
\textbf{\tool{} as a complementary resource.}
Rather than replacing current directories, we developed \tool{} as a tool to complement these existing directories. 
\tool{} can be used in conjunction with the directories of universities, companies, agencies, and other institutions to lend an additional perspective upon the diversity of research interests that the institution holds. 
\medskip

\section{Conclusion and Future Work}
\label{sec:conclusion}
    
    \tool{}, in its current form, will continue to be useful for years to come, but we plan on continuing to improve the system by increasing the sophistication of the NLP techniques used in analysis and expanding the available functionalities for exploring researcher datasets. In the current version of \tool{}, we use TFIDF to generate researcher embeddings (Section \ref{sec:wordembedding}) from our gathered researcher data before using PCA and Gaussian mixture modeling for visualizing these embeddings and performing clustering techniques (Section \ref{sec:clusteringtechniques}). However, as we seek to increase the complexity of our embeddings, we plan on exploring several potential embedding techniques. For example, we aim to extract hidden layers from pretrained and finetuned Transformer \cite{vaswani2017attention} models such as BERT \cite{devlin2018bert}. Prior work has explored fine-tuning these models on text data from the scientific domain, yielding improved results on downstream tasks \cite{beltagy2019scibert}. However, we aim to use similar techniques in the context of visualization. Using these techniques, we open up the possibility of both improved information extraction \textit{and} visualization of researcher datasets.
    
    Lastly, we hope that \tool{} can assist any individual seeking to delve deeper into the fields of interests found within any group of researchers. 
    We encourage any institution composed of published researchers to use \tool{} if they would like to explore the diversity of content produced by their members. We expect that recommendation systems for research papers and publication venues will continue to be a topic of interest in coming years, as there have been several different studies addressing potential platforms and solutions \cite{beel2016paper, medvet2014publication, beel2017towards, alhoori2017recommendation, kuccuktuncc2013theadvisor}. Furthermore, we also expect organizations will seek to improve outdated directory systems so that both internal and external groups can more efficiently and confidently connect with researchers for potential collaborations.

\balance

\bibliographystyle{ACM-Reference-Format}
\bibliography{cikm2020}

\end{document}